\documentclass[preprint,aps,pra,showpacs,floatfix]{revtex4}
\usepackage{graphicx}
\usepackage{times}
\usepackage{nicefrac}
\usepackage{amsmath}
\usepackage{amsfonts}
\usepackage{amssymb}
\usepackage{amsthm}
\usepackage{epsf}
\usepackage{bm}
\usepackage{bbm}

\usepackage{dcolumn}
\newcolumntype{.}{D{x}{}{-1}}

\newcommand{\be}{\begin{eqnarray}}
\newcommand{\ee}{\end{eqnarray}}
\newcommand{\la}{\langle}
\newcommand{\ra}{\rangle}

\newcommand{\veps}{\varepsilon}

\newcommand{\pr}{\prime}
\newcommand{\hsp}{\hspace}
\newcommand{\balpha}{\bm{\alpha}}

\newcommand{\bmu}{\bm{\mu}}
\newcommand{\bfT}{{\bf T}}
\newcommand{\rmd}{{\rm d}}

\newcommand{\bfn}{{\bf n}}

\newcommand{\rpr}{r^\prime}

\newcommand{\bfr}{{\bf r}}

\newcommand{\ddd}{\rmd^3}

\newcommand{\aZ}{\alpha Z}

\begin{document}

\title{Ground-state hyperfine structure of H-, Li-, and B-like ions
in middle-$Z$ region}
\author{A. V. Volotka,$^{1,2}$ D. A. Glazov,$^{2}$ I. I. Tupitsyn,$^{2}$
N. S. Oreshkina,$^2$ G. Plunien,$^{1}$ and V. M. Shabaev$^{2}$}

\affiliation{
$^1$ Institut f\"ur Theoretische Physik, Technische Universit\"at Dresden,
Mommsenstra{\ss}e 13, D-01062 Dresden, Germany \\
$^2$ Department of Physics, St. Petersburg State University,
Oulianovskaya 1, Petrodvorets, 198504 St. Petersburg, Russia \\
}

\begin{abstract}
The hyperfine splitting of the ground state of H-, Li-, and B-like ions
is investigated in details within the range of nuclear numbers $Z=7-28$.
The rigorous QED approach together with
the large-scale configuration-interaction Dirac-Fock-Sturm
method are employed for the evaluation of the
interelectronic-interaction contributions
of first and higher orders in $1/Z$.
The screened QED corrections are evaluated to all orders in
$\aZ$ utilizing an effective potential approach.
The influence of nuclear magnetization distribution
is taken into account
within the single-particle nuclear model.
The specific differences between the hyperfine-structure level shifts
of H- and Li-like ions, where the uncertainties associated with
the nuclear structure corrections are significantly reduced, 
are also calculated.
\end{abstract}


\maketitle

%%%%%%%%%%%%%%%%%%%%%%%%%%%%%%%%%%%%%%%%%%%%%%%%%%%%%%%%%%%%%%%%%%%%%%%%
%
%%%%%%%%%%%%%%%%%%%%%%%%%%%%%%%%%%%%%%%%%%%%%%%%%%%%%%%%%%%%%%%%%%%%%%%%
%
\section{Introduction}

Accurate knowledge of the hyperfine structure lines of middle-$Z$
multicharged ions is of great interest due to suggested observations
of these lines from hot rarefied astrophysical plasmas
\cite{sunyaev:1984:483,sunyaev:2007:83}.
Such observations may allow one to study the chemical and isotopic
compositions of the supernova remnants, and the hot interstellar
medium, including the galactic halos, which are the main types of
objects from which intense emission lines are expected.
The experiments on the determination of hyperfine splittings
will also enable us to refine the deduction of nuclear magnetic
moments of different isotopes and to inspect the various computational
models employed for the theoretical description of nuclear effects.
High-precision measurements of the ground-state hyperfine structure
of heavy highly charged ions have been performed in
Refs.~\cite{klaft:1994:2425,crespo:1996:826,crespo:1998:879,
seelig:1998:4824,beiersdorfer:2001:032506}.
Extension of these experiments to Li-like ions presently being 
prepared \cite{winters:2007:403} will provide tests of quantum
electrodynamics (QED) in strong electric and magnetic fields
on level of a few percent in specific difference of the hyperfine
splitting values of H- and Li-like ions
\cite{shabaev:2001:3959}.
In this difference the main theoretical uncertainty
which originates from the nuclear magnetization
distribution correction (Bohr-Weisskopf effect)
is essentially reduced.
In specific differences of heavy H- and B-like ions or
Li- and B-like ions the same reduction of the theoretical
uncertainty can be also achieved. This becomes clear from the approximate
analytical expressions for the Bohr-Weisskopf correction given
in Ref.~\cite{shabaev:1994:5825}.

The theoretical investigations of the hyperfine splitting
of H- and Li-like multicharged ions
in the middle-$Z$ region have some history.
The first accurate calculation ($\sim 0.1\%$), based on a combination
of $1/Z$ perturbation theory and the nonrelativistic
configuration-interaction Hartree-Fock method, was performed
in Refs.~\cite{shabaev:1995:3686,shabaev:1997:243}.
Later, Boucard and Indelicato \cite{boucard:2000:59}
employing the multi-configuration Dirac-Fock method
presented the evaluation of the hyperfine splitting values
over the entire range of the nuclear charge numbers $Z=3-92$.
Expansion in $\alpha Z$ of the QED correction has been worked out
in Refs.~\cite{pachucki:1996:1994,nio:1997:7267,karshenboim:2002:13}
(for earlier studies  see references therein and recent
reviews \cite{mohr:2005:1,karshenboim:2005:1}).
However, the application of the $\alpha Z$-expansion
is restricted to $s$-states in one-electron ions
and limited by its convergence property.
Therefore, we evaluate the radiative corrections
numerically to all orders in $\aZ$ accounting for the
interelectronic-interaction effects by means of local screening 
potentials.
All-order calculations of one-loop
QED contributions to the hyperfine structure
for middle-$Z$ ions have been previously performed
for the $1s$ state \cite{blundell:1997:4914,sunnergren:1998:1055,
yerokhin:2001:012506,yerokhin:2005:052510},
for the $2s$ state \cite{yerokhin:2001:012506,yerokhin:2005:052510},
and for the $2p_{1/2}$ state \cite{sapirstein:2006:042513}.
However, almost all these calculations of the QED corrections
were dealing with one-electron ions only,
where the screening effects are absent.

In the present paper, we calculate the ground-state
hyperfine structure of H-, Li-, and B-like sequences
in the middle-$Z$ region. The one-loop radiative corrections
are evaluated to all orders in $\aZ$ employing an effective
local screening potential. Many-body effects are taken
into account to the first order in $1/Z$ within the QED perturbation
theory and to higher orders within the large-scale configuration-interaction
Dirac-Fock-Sturm method (CI-DFS).
The single-particle nuclear model is employed for
the evaluation of the Bohr-Weisskopf correction.
The main goal of this work is to improve the accuracy of previous
results for the hyperfine structure of H- and Li-like ions
and to present novel calculations for the B-like sequence
in the middle-$Z$ region.

The paper is organized as follows: In the next section
the basic formulas for the hyperfine splitting are given
and the derivation of the various contributions is described.
In Section~\ref{sec-3} we present the numerical results
for all contributions and compare the total values with previously
reported calculations and with existing experimental data.
Section~\ref{sec-4} provides a complete compilation of the total values
for the hyperfine splitting of H-, Li, and B-like ions
as well as the results for the specific differences
between the hyperfine structure of H- and Li-like ions.
We close with a short summary and point out the main achievements of
the present work.

Relativistic units ($\hbar = 1$, $c = 1$, $m = 1$) and the Heaviside
charge unit [$\alpha = e^2/(4\pi)$, $e<0$] are used throughout
the paper.
%
%%%%%%%%%%%%%%%%%%%%%%%%%%%%%%%%%%%%%%%%%%%%%%%%%%%%%%%%%%%%%%%%%%%%%%%%
%
\section{Basic expressions}
\label{sec-2}
The interaction of atomic electrons with the nuclear magnetic-dipole
moment is described by the Fermi-Breit operator, which is conveniently
written as a scalar product of two tensor operators
\be
  H_\mu = \frac{|e|}{4\pi}\,\bmu \cdot \bfT\,,
\ee
where $\bmu$ is the nuclear magnetic moment operator acting in the space of nuclear states.
The electron part $\bfT$ is defined by the following expression
\be
\label{bfT}  
  \bfT = \sum_i \frac{[\bfn_i\times\balpha_i]}{r_i^2}\,,
\ee
where index $i$ refers to the $i$-th electron of the atom,
$\balpha$ is the Dirac-matrix vector, and $\bfn_i = \bfr_i/r_i$.
This interaction leads to the hyperfine splitting of the atomic levels.
For an ion with one electron (e.g., $ns$ or $np_{1/2}$ state) over the closed
shells this splitting can be written in the form
\be
\label{hfs:1}
  \Delta E^{(a)} &=& \frac{\alpha(\aZ)^3}{n^3} \frac{g_I}{m_p}
    \frac{2I+1}{(j+1)(2l+1)}\frac{1}{(1+\frac{m}{M})^3}\nonumber\\
  &&\times\left[ A(\aZ)(1-\delta)(1-\veps) + \frac{1}{Z}B(\aZ)
    + \frac{1}{Z^2}C(Z,\aZ) + x_{\rm{rad}}\right]\,.
\ee
Here $Z$ is the nuclear charge number, $m_p$ and $M$
are the proton and nuclear masses, respectively.
Within the approximation of noninteracting electrons,
where the contribution of the closed shells is neglected,
the hyperfine splitting is explicitely determined by the quantum numbers
of valence electron state $a$, which is characterized by the principal quantum
number $n$, the angular momentum $j$, its projection $m_j$, and the parity $l$.
A nucleus with spin $I$ possesses a nuclear $g$ factor $g_I=\mu/\mu_N I$,
where $\mu$ is the nuclear magnetic moment and $\mu_N$ is the
nuclear magneton.
$A(\alpha Z)$ is the one-electron relativistic factor, $\delta$
and $\veps$ are, respectively, the corrections for distributions
of the charge and magnetic moment over the nucleus; the functions
$B(\alpha Z)$ and $C(Z,\alpha Z)$ determine the corrections for
the electron-electron interaction of first and higher orders
in $1/Z$, respectively; $x_{\rm rad}$ is the QED correction.
These terms are subsequently described in the following
subsections.
\subsection{One-electron contributions}
\label{subsection2A}
The relativistic factor $A(\aZ)$ corresponding to the point-like nucleus 
is known analytically \cite{pyykko:1973:785}
\be
\label{A}
  A(\aZ) = \frac{n^3(2l+1)\kappa[2\kappa(\gamma + n_r) - N]}
           {N^4\gamma(4\gamma^2 - 1)}\,,
\ee
where $n_r = n - |\kappa|$ is the radial quantum number,
$\kappa = (-1)^{j+l+1/2}(j+1/2)$, $\gamma = \sqrt{\kappa^2-(\aZ)^2}$,
$N = \sqrt{n_r^2 + 2n_r\gamma + \kappa^2}$.
The nuclear charge distribution correction $\delta$
can be found either analytically \cite{shabaev:1994:5825,volotka:2003:51}
or numerically by solving the Dirac equation with the Coulomb potential of the
extended nucleus. In this work it is evaluated numerically
employing the homogeneously-charged-sphere model for the nuclear
charge distribution.
In order to estimate the uncertainty due to the model dependence 
the Fermi model is used as well.
The Bohr-Weisskopf correction $\veps$ originates from
the spatial distribution of the magnetic moment inside the nucleus.
For a rigorous treatment of this effect for low-$Z$ systems we refer
to Ref.~\cite{pachucki:2007:022508}. In the present work
we restrict our consideration to models in which
it can be accounted for by replacing the factor $1/r^2$
in Eq.~(\ref{bfT}) by $F(r)/r^2$,
where $F(r)$ is the volume distribution function.
For example, in case of the sphere model it reads
\be
\label{BW:2}
  F(r) = \left\{\begin{array}{cr}
     \displaystyle\left(\frac{r}{R_0}\right)^3, & r \le R_0\\
     \displaystyle                           1, & r > R_0
                \end{array}\right.\,,
\ee
where $R_0 = \sqrt{5/3}\,\la r^2\ra^{1/2}$ is the radius of the sphere,
and $\la r^2\ra^{1/2}$ is the charge root-mean-square radius of the nucleus.
However, with the sphere model one can not always describe adequately
the nuclear magnetization distribution.
The approximation of the nuclear single-particle model is widely used
for the evaluation of the Bohr-Weisskopf correction
\cite{shabaev:1994:5825,shabaev:1995:3686,shabaev:1997:243,
shabaev:1997:252,shabaev:1998:149,zherebtsov:2000:701,tupitsyn:2002:389}.
Within this model the nuclear magnetization is determined
by the total angular momentum of the unpaired nucleon (proton or neutron).
Accordingly, the nuclear $g$ factor $g_I$ is just the
Land\'e factor of an extra nucleon, which is defined by
the well-known formula
\be
\label{g_I}
  g_I = \mu/\mu_N I =
        \frac{1}{2}\left[(g_L+g_S) + (g_L-g_S) \frac{L(L + 1)-3/4}{I(I + 1)}\right]\,,
\ee
where $L$ is the nuclear orbital momentum, $g_L$ and $g_S$ are the
orbital and spin $g$ factors of the valence nucleon, respectively.
In case of a valence proton $g_L = 1$, while
for an extra neutron $g_L = 0$;
$g_S$ is chosen such as to reproduce the experimental value of the nuclear
magnetic moment $\mu$ according to Eq.~(\ref{g_I}).
For nuclei with odd or even nuclear charge numbers
the role of the unpaired nucleon is either played by a proton
or a neutron, respectively.
In the framework of the nuclear single-particle
model the radially symmetric distribution function $F(r)$ has been derived
in Refs.~\cite{shabaev:1997:252,zherebtsov:2000:701,tupitsyn:2002:389}.
Here we neglect the contribution of the spin-orbit interaction
and employ the homogeneous distribution
for the radial part of the odd nucleon wavefunction
inside the nucleus \cite{shabaev:1994:5825,tupitsyn:2002:389}.
In this approximation $F(r)$ reads
\be
\label{BW:3}
  F(r) = \left(\frac{r}{R_0}\right)^3
         \left\{ 1 - 3 \ln{\left(\frac{r}{R_0}\right)}\,\frac{\mu_N}{\mu}
         \left[-\frac{2I-1}{8(I+1)} g_S + \left(I-\frac{1}{2}\right) g_L\right]\right\}\,,
         \hsp{5mm} r\le R_0\,,
\ee
for $I=L+\frac{1}{2}$ and
\be
\label{BW:4}
  F(r) = \left(\frac{r}{R_0}\right)^3
         \left\{ 1 - 3 \ln{\left(\frac{r}{R_0}\right)}\,\frac{\mu_N}{\mu}
         \left[\frac{2I+3}{8(I+1)} g_S + \frac{I(2I+3)}{2(I+1)} g_L\right]\right\}\,,
         \hsp{5mm} r\le R_0\,,
\ee
for $I=L-\frac{1}{2}$. For $r>R_0$ the distribution function $F(r) = 1$.
In the case of $^{14}$N with $I = 1$ we follow the work \cite{shabaev:1995:3686}
and assume that the nuclear magnetization is determined by
the odd proton and neutron. The corresponding formulas for $\veps$ were derived
in Refs.~\cite{bellac:1963:645,shabaev:1995:3686}.
The uncertainty of the Bohr-Weisskopf correction is estimated
as the maximum of two values:
50\% of $\veps$ itself and the difference between
$\veps$ obtained in single-particle and sphere nuclear models.
As in our previous studies (see, e.g., the related discussion
in Ref.~\cite{shabaev:1997:252}), the uncertainty obtained by this
procedure must generally be considered only as the order of 
magnitude of the expected error bar. More accurate calculations of 
the Bohr-Weisskopf effect must be based on many-particle nuclear models 
and should include a more rigorous procedure for determination 
of the uncertainty.
The nuclear vector polarizability correction derived in Ref.~\cite{pachucki:2007:022508}
is assumed to contribute less than the uncertainty of $\veps$ indicated above.

Separating out the nuclear parameters and the nonrelativistic value
of the hyperfine splitting one finds that
the one-electron contributions considered above can be numerically evaluated
in terms of a matrix element
\be
  A(\aZ)(1-\delta)(1-\veps) = G_a \, \la a | T_0 | a \ra
\ee
of the zero component $T_0$ of the operator $\bfT$
given by Eq.~(\ref{bfT}), multiplied
by the magnetization distribution function $F(r)$.
The wavefunction $| a \ra$ of the valence state,
characterized by quantum numbers $a$ $=$ $n$, $j$, $m_j$, and $l$,
is obtained as a solution of the Dirac equation
with the potential of the extended nucleus.
The multiplicative factor $G_a$ reads 
\be
  G_a = \frac{n^3 (2l+1) j (j+1)}{2 (\aZ)^3 m_j}\,.
\ee
\subsection{Many-electron contributions}
Now we pass to the many-electron corrections.
The term $B(\aZ)/Z$ in Eq.~(\ref{hfs:1}) determines the interelectronic-interaction
correction of the first order in $1/Z$. A rigorous QED treatment
of this contribution can be carried out utilizing the
two-time Green's function method \cite{shabaev:2002:119}.
To simplify the derivation of formal expressions, it is
convenient to incorporate the core
electrons as belonging to a redefined vacuum. This leads
to merging the interelectronic-interaction correction of
order $1/Z$ with the one-loop radiative corrections.
Such a treatment was applied previously in
Refs.~\cite{shabaeva:1992:72,shabaeva:1995:2811}.
The corresponding
expression for the interelectronic-interaction
correction reads
\be
\label{B(aZ)}
  B(\aZ)/Z  = 
  2\,G_a
\sum_c &\Biggl\{& \sum^{\veps_n \not = \veps_a}_n
    \frac{ \la a c | I(0) | n c \ra \la n | T_0 | a \ra }
    {\veps_a-\veps_n} + \sum^{\veps_n \not = \veps_c}_n
    \frac{ \la a c | I(0) | a n \ra
    \la n | T_0 | c \ra } {\veps_c-\veps_n}
\nonumber\\
  &&
    - \sum^{\veps_n \not = \veps_a}_n \frac{ \la a c | I(\veps_a-\veps_c)
    | c n \ra \la n | T_0 | a \ra } {\veps_a-\veps_n}
    -\sum^{\veps_n \not = \veps_c}_n \frac{ \la a c | I(\veps_a-\veps_c) | n a \ra
    \la n | T_0 | c \ra } {\veps_c-\veps_n}
\nonumber\\
  && -
    \frac{1}{2} \left[ \la a | T_0 | a \ra - \la c | T_0 | c \ra \right]
    \la a c | I^{\prime}(\veps_a-\veps_c) | c a \ra
    \Biggr\} \,,
\ee
where $\veps_m$ are the one-electron energies,
$I(\omega) =  e^2 \alpha^{\mu} \alpha^{\nu} D_{\mu\nu}(\omega)$,
$I^{\prime}(\omega) = \rmd I(\omega) / \rmd \omega$, 
$\alpha^{\mu}=(1,\balpha)$,
and $D_{\mu\nu}(\omega)$ is the photon propagator.
It should be noted that the total $1/Z$ interelectronic-interaction
correction given by Eq.~(\ref{B(aZ)}) is gauge independent.
We perform the calculation employing Coulomb and Feynman gauges for the
photon propagator, thus receiving an accurate check of
the gauge invariance of the results.

The interelectronic-interaction correction of higher orders
$C(Z,\aZ)/Z^2$ is calculated within the framework 
of the large-scale configuration-interaction method
in the basis of Dirac-Fock-Sturm orbitals
\cite{bratsev:1977:2655}.
This method was successfully employed in our previous atomic calculations
\cite{tupitsyn:2003:022511,glazov:2004:062104,
shabaev:2005:062105,tupitsyn:2005:062503,artemyev:2007:173004}.
The interelectronic-interaction operator employed
in the Dirac-Coulomb-Breit equation reads
\be
\label{int}
  V_{\rm int}  = \lambda\alpha
  \sum_{i<j}\left\{ \frac{1}{r_{ij}}
                -\frac{\balpha_i \cdot \balpha_j}{2r_{ij}}
                -\frac{(\balpha_i \cdot \bfr_i)(\balpha_j \cdot \bfr_j)}{2r^3_{ij}}
          \right\}\,,
\ee
where the sum runs over all electrons. A scaling parameter
$\lambda$ is introduced to separate
terms of different order in $1/Z$ from the numerical results
with different $\lambda$. This representation allows us to perform
the expansion in powers of $\lambda$.
In this way, the higher-order term is written as
\be
\label{C(aZ)}
  C(Z,\aZ)/Z^2  &=& G_a \Biggl\{
    \la \Psi_\lambda(\gamma JM_J) |
     T_0 | \Psi_\lambda(\gamma JM_J) \ra \Bigr|_{\lambda=1}
  - \la \Psi_\lambda(\gamma JM_J) | 
     T_0 | \Psi_\lambda(\gamma JM_J) \ra \Bigr|_{\lambda=0}
  \nonumber\\
&&- \frac{\rmd}{\rmd\lambda} \la \Psi_\lambda(\gamma JM_J) |
     T_0 | \Psi_\lambda(\gamma JM_J) \ra \Bigr|_{\lambda=0}
\Biggl\}\,.
\ee
The many-electron wavefunction $\Psi_\lambda(\gamma JM_J)$
is characterized by the total angular momentum $J$, its projection $M_J$, and
the rest quantum numbers $\gamma$.
The configuration-interaction matrix contains all single,
double, and triple positive-energy excitations.
Single-electron excitations to the negative-energy spectrum
were accounted for in the many-electron wavefunction
$\Psi_\lambda(\gamma JM_J)$
employing perturbation theory.

The calculation of the interelectronic-interaction corrections
$B(\aZ)/Z$ and $C(Z,\aZ)/Z^2$ is performed employing
the homogeneously-charged-sphere model for the nuclear
charge distribution and single-particle model for the nuclear
magnetic moment distribution.
\subsection{One-loop radiative contribution}
\begin{figure}
\includegraphics{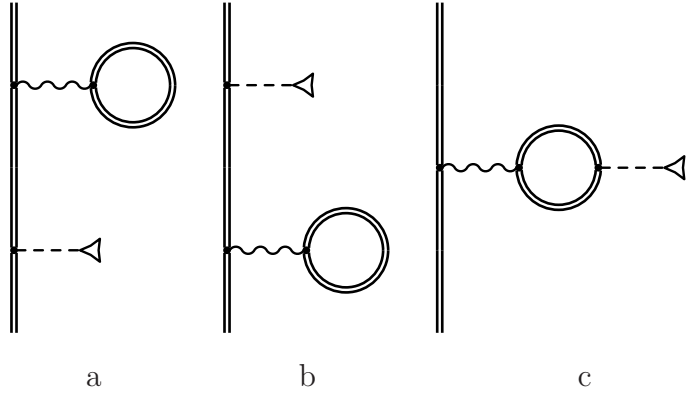} \caption {Feynman diagrams
representing the vacuum-polarization correction to the hyperfine
splitting. The wavy line indicates the photon propagator
and the double line indicates the bound-electron wavefunctions
and propagators.
The dashed line terminated with the triangle denotes the
hyperfine interaction.}
\label{VP}
\end{figure}
\begin{figure}
\includegraphics{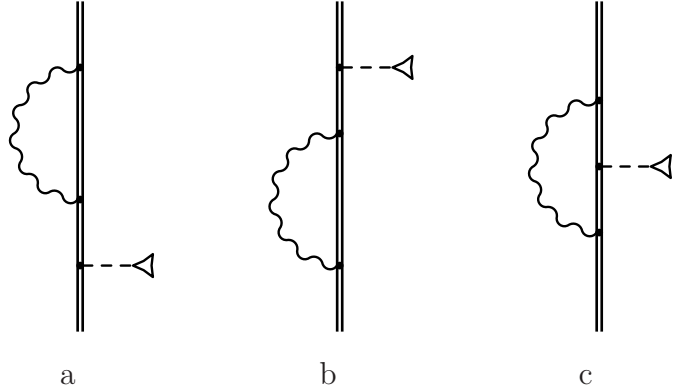} \caption {Feynman diagrams
representing the self-energy correction to the hyperfine
splitting. Notations are the same as in Fig.~\ref{VP}.}
\label{SE}
\end{figure}
The one-loop radiative contribution $x_{\rm rad}$
appears as the sum of vacuum-polarization (VP)
and self-energy (SE) corrections, $x_{\rm rad} = x_{\rm VP} + x_{\rm SE}$,
as depicted diagrammatically in Figs.~\ref{VP} and \ref{SE},
respectively.
However, for the Li- and B-like ions
along with the one-electron part the correction $x_{\rm rad}$ contains 
also the many-electron part.
In order to account for many-electron effects
we consider an effective spherically symmetric potential
$V_{\rm eff}$ that partly takes into account the
interelectronic interaction
between the valence electron $a$ and the core electrons $c$.
This can be achieved by means of the Kohn-Sham screening potential
derived within the density-functional theory \cite{kohn:1965:A1133}
\be
\label{Veff}
  V_{\rm eff}(r) = V_{\rm nuc}(r)
   + \alpha \int_0^\infty \rmd r'\frac{1}{r_>}\rho_t(r')
   - \frac{2}{3}\frac{\alpha}{r}\left(
      \frac{81}{32\pi^2}r\rho_t(r)\right)^{1/3}\,,
\ee
which we employed successfully in previous calculations
\cite{glazov:2006:330,oreshkina:2007:889,kozhedub:2007:012511,oreshkina:2008:675}.
Here $V_{\rm nuc}$ is the potential of the extended nucleus
and $\rho_t$ denotes the total one-electron density.
In order to estimate the sensitivity of the result
on the specific choice of the screening potential we consider also
the core-Hartree potential.

The VP correction $x_{\rm VP}$ is divided into the electric-loop
part, Fig.~\ref{VP} (a,b), which accounts for the VP correction to the scalar
binding potential $V_{\rm eff}$, and the magnetic-loop
part, Fig.~\ref{VP} (c), corresponding to the VP-corrected hyperfine
interaction potential.
The expression for the electric-loop term  Fig.~\ref{VP} (a,b) reads
\be
  x_{\rm VP}^{\rm el} = 2\,G_a \sum_n^{\veps_n\neq\veps_a}
   \frac{\la a | T_0 | n\ra \la n | U_{\rm VP}^{\rm el} | a \ra}
   {\veps_a-\veps_n}
   \,,
\ee
where $U_{\rm VP}^{\rm el}$ represents the renormalized one-loop
VP potential. 
It is divided into the Uehling and Wichmann-Kroll parts,
$U_{\rm VP}^{\rm el} = U_{\rm VP}^{\rm Ue-el} + U_{\rm VP}^{\rm WK-el}$.
The Uehling part can be evaluated according to the well-known
equation
\be
  U_{\rm VP}^{\rm Ue-el}(r) = -\frac{2\alpha^2 Z}{3\pi}\int_1^\infty
  \rmd t \, \frac{\sqrt{t^2-1}}{t^2} \left( 1+\frac{1}{2t^2} \right)
  \int \ddd \rpr \, \frac{\rho_{\rm eff}(\bfr^\pr)}{|\bfr-\bfr^\pr|}\,
  e^{-2|\bfr-\bfr^\pr|t}\,,
\ee
where the density $\rho_{\rm eff}$ is related to the effective binding
potential $V_{\rm eff}$ $[\,$via the Poisson equation
$\,\Delta V_{\rm eff}(\bfr) = 4\pi\alpha Z\,\rho_{\rm eff}(\bfr)\,]$.
The Wichmann-Kroll part can be generated by summing
up the partial-wave differences between the unrenormalized
total VP potential and the unrenormalized Uehling term
\cite{soff:1988:5066,manakov:1989:1167}.
In this work we employ the approximate formula for the Wichmann-Kroll
electric-loop potential derived in Ref.~\cite{fainshtein:1990:559}.
The correction to the hyperfine splitting due to the
magnetic loop $x_{\rm VP}^{\rm ml}$
can be written in the form
\be
  x_{\rm VP}^{\rm ml} = G_a\,\la a | U_{\rm VP}^{\rm ml} | a \ra\,,
\ee
where $U_{\rm VP}^{\rm ml}$ is the VP-corrected hyperfine potential 
$T_0$. It can be renormalized utilizing the same scheme
as for the electric loop.
For the distribution function $F(r)$ corresponding
to the sphere model (see Eq.~(\ref{BW:2})) we obtain the
following analytical expression for the magnetic-loop
Uehling term
\be
  U_{\rm VP}^{\rm Ue-ml}(r) &=& \frac{\alpha}{\pi}
   \frac{[\bfn\times\balpha]_0}{r^2}\,\frac{3}{16R_0^3}
     \Biggl\{ 4rR_0\Bigl[\beta_1(R_0+r)+\beta_1(|R_0-r|)\Bigr]
            +2(R_0+r)\,\beta_2(R_0+r)\nonumber\\
          &&-2|R_0-r|\,\beta_2(|R_0-r|)
            +\beta_3(R_0+r)-\beta_3(|R_0-r|)\Biggr\}\,,
\ee
where the function $\beta_n$ is defined as
\be
  \beta_n(r) = \frac{2}{3}\int_1^\infty
   \rmd t \frac{\sqrt{t^2-1}}{t^{n+2}}\,
   \left(1+\frac{1}{2t^2}\right) e^{-2tr}\,.
\ee
The contribution of the remaining Wichmann-Kroll
magnetic-loop term is relatively small.
For the case of the $1s$ and $2s$ states the values for the
Wichmann-Kroll magnetic-loop term are taken from
Ref.~\cite{yerokhin:2005:052510}.
For $2s$ we have also incorporated the
screening effect, assuming that the screening coefficient
is the same as for the Uehling magnetic-loop term.
For the $2p_{1/2}$ state $x_{\rm VP}^{\rm WK-ml}$
turns out to be smaller than the uncertainties assigned to the
calculation.

Now let us turn to the evaluation of the SE correction.
The formal expression can be derived by means
of the two-time Green's function method \cite{shabaev:2002:119}.
The SE correction appears as the sum,
$x_{\rm SE} = x_{\rm SE}^{\rm irr} + x_{\rm SE}^{\rm red} + x_{\rm SE}^{\rm ver}$,
of the irreducible $x_{\rm SE}^{\rm irr}$, reducible
$x_{\rm SE}^{\rm red}$, and
vertex $x_{\rm SE}^{\rm ver}$ terms, respectively.
The irreducible part, depicted in Fig.~\ref{SE} (a,b) with
the intermediate state energy $\veps_n\neq\veps_a$, is represented by the expression
\be
  x_{\rm SE}^{\rm irr} = 2 G_a \sum_n^{\veps_n\neq\veps_a}
   \frac{\la a | T_0 | n\ra \la n |
   \Bigl[\Sigma(\veps_a)-\gamma^0\delta m\Bigr] | a \ra}{\veps_a-\veps_n}
   \,,
\ee
where $\delta m$ is the mass counter-term and
$\Sigma(\veps)$ denotes the unrenormalized
self-energy operator with matrix elements defined by
\be
  \la a | \Sigma(\veps) | b \ra =
   \frac{i}{2\pi} \int_{-\infty}^\infty
   \rmd \omega \sum_n \frac{\la a n | I(\omega) | n b \ra}
   {\veps - \omega - \veps_n ( 1 - i0 )}
   \,.
\ee
Accordingly, the irreducible contribution
can be written as nondiagonal matrix element of
the self-energy operator. Thus, the renormalization
scheme developed for the first-order self-energy
correction can be also applied in this case (see, e.g.,
Refs.~\cite{mohr:1974:26,snyderman:1991:43,yerokhin:1999:800}).
Only a slight extension of the corresponding formulas for the
case of a nondiagonal matrix element is needed.

The expression for the reducible term is given by
\be
\label{red}
  x_{\rm SE}^{\rm red} = G_a \la a|
  \frac{\rmd \Sigma(\veps)}{\rmd\veps} \Bigl|_{\veps=\veps_a}|a\ra
  \la a|T_0|a\ra\,,
\ee
while the vertex part, Fig.~\ref{SE} (c), reads
\be
\label{ver}
  x_{\rm SE}^{\rm ver} = G_a
  \frac{i}{2\pi} \int^\infty_{-\infty}\rmd \omega \sum_{n_{1} \, n_2}
  \frac{\la a n_2|I(\omega)|n_1 a\ra \, \la n_1|T_0|n_2\ra}
  {(\veps_a-\omega-\veps_{n_1}(1-i0))(\veps_a-\omega-\veps_{n_2}(1-i0))}\,.
\ee
Both reducible and vertex terms are ultraviolet-divergent.
In order to isolate the divergencies in a covariant way,
we separate out the zero-potential terms in which bound
electron propagators are
replaced by free propagators. The sum of the latter terms
for the reducible and the vertex part is denoted by
$x_{\rm SE}^{\rm vr(0)} = x_{\rm SE}^{\rm red(0)} + x_{\rm SE}^{\rm ver(0)}$.
Their evaluation is performed in momentum space, where the ultraviolet
divergencies can be canceled in a standard way.
The remaining part of the reducible and the vertex
contribution $x_{\rm SE}^{\rm vr(1+)}$ is ultraviolet finite.
However, we note that the term with
$\veps_{n_1} = \veps_{n_2} = \veps_{a}$ in
Eq.~(\ref{ver}) involves an infrared divergency, which is canceled
by the corresponding term of the reducible contribution.
Performing the integration over the energy
of the virtual photon in these terms analytically
we explicitly achieve the finite result in the sum
of the reducible and vertex contributions.
The remaining part of the many-potential term
$x_{\rm SE}^{\rm vr(1+)}$ is infrared finite,
and can be calculated in coordinate space by means of a point-by-point
subtraction of the corresponding contributions with
free propagators inside of the self-energy loop.
Angular integration and summation over intermediate angular
momentum projections is carried out in a standard way \cite{yerokhin:1999:800}.
The evaluation of the many-potential terms performed in the coordinate
space involves an infinite summation
over the angular-momentum quantum number $\kappa$
of intermediate states. This sum
was extended up to $|\kappa_{\rm max}| = 10$
and the remaining part of the sum is estimated by a least-square
inverse-polynomial fitting.
One also observes that the results
of the radial integration converge better,
when an extended model for the magnetization
distribution is employed.
In our calculations of the radiative corrections
we have utilized the sphere model for the magnetic moment
distribution function $F(r)$.
%
%%%%%%%%%%%%%%%%%%%%%%%%%%%%%%%%%%%%%%%%%%%%%%%%%%%%%%%%%%%%%%%%%%%%%%%%
%
\section{Numerical results}
\label{sec-3}
Now let us pass to the presentation of the numerical procedure and
the results for H-, Li-, and B-like sequences.
The infinite summations over the complete spectrum
of the Dirac equation involved in the numerical evaluations are
performed employing the
finite-basis set approach.
The B-splines basis set was constructed
utilizing the dual kinetic balance approach
\cite{shabaev:2004:130405}.
The latter treats large and small components on equal
footing and respects the charge conjugation symmetry.
As a consequence no unphysical spurious states
appear and moreover, it improves the convergence
properties and the accuracy considerably.
The values of the nuclear root-mean-square radii are
taken from the tabulation \cite{angeli:2004:185}.
The root-mean-square radius, in particular, for the $^{33}$S
isotope
is assumed to be the average of the values given for even isotopes
$^{32}$S
and $^{34}$S, respectively.
Empirical data for the nuclear properties:
spin $I$, parity $\pi$, and magnetic moment $\mu/\mu_N$
are taken from Ref.~\cite{stone:2005:75}.
All these values are also compiled in Table~\ref{tab:H}.
We indicate the uncertainties assigned to the nuclear magnetic moments
only if they exceed the level of $10^{-5}$ in the relative
units. One has to note here, that the magnetic moment
values obtained via the nuclear magnetic resonance technique
do not usually account for the chemical shift
\cite{gustavsson:1998:3611},
which is of the order $10^{-3}-10^{-4}$ or sometimes even larger.
\subsection{H-like ions}
The individual contributions to the hyperfine splitting
for the light H-like ions are presented in Table~\ref{tab:H}.
The values obtained for the radiative corrections $x_{\rm SE}$ and $x_{\rm VP}$
are in good agreement with the most accurate
nonperturbative results \cite{yerokhin:2005:052510}
based on the Coulomb-Dirac Green function.
The slight difference is explained by the finite-nuclear-size effects
accounted for in the present work.
The latter is especially important
for the evaluation of the specific difference between H- and 
Li-like hyperfine splitting values, where the QED corrections
have to be calculated within the same nuclear model.
As one can see from the table the main uncertainty
originates from the Bohr-Weisskopf correction $\veps$.
In Table~\ref{tab:H:comp}
the predictions for the total transition energies
$\Delta E^{(1s)}$
are compared with the results of previous calculations
\cite{shabaev:1995:3686,boucard:2000:59}.
Deviations between our results and those reported
in Ref.~\cite{boucard:2000:59}
arise from the different treatment of the Bohr-Weisskopf
effect.
In work \cite{boucard:2000:59} the simple spherical
model for the magnetization distribution was employed.
Theoretical values
for the total transition energies $\Delta E^{(1s)}$ and wavelengths $\lambda^{(1s)}$
are presented
in Table~\ref{tab:total} below.
\begin{table}
\caption{Individual contributions to the ground-state hyperfine splitting
of the hydrogenlike ions.}
\label{tab:H}
\tabcolsep10pt
\scriptsize
\begin{tabular}{lcllccccc} \hline
Ion               &  $I^\pi$        & $\mu/\mu_N$    & $\la r^2 \ra^{1/2}$
     & $A(\aZ)$ & $\delta$    & $\veps$      & $x_{\rm SE}$ & $x_{\rm VP}$ \\ \hline
$^{14}$N$^{6+}$   &  $1+$           &  0.40376       &  2.5579  
     &  1.00393 &  0.00067(3) & -0.00004(28) &  0.00009     &  0.00028     \\
$^{15}$N$^{6+}$   &  $\frac{1}{2}-$ & -0.28319       &  2.6061  
     &  1.00393 &  0.00068(3) &  0.00114(88) &  0.00009     &  0.00028     \\
$^{17}$O$^{7+}$   &  $\frac{5}{2}+$ & -1.8938(1)     &  2.6953  
     &  1.00514 &  0.00081(3) &  0.00033(17) & -0.00006     &  0.00032     \\
$^{19}$F$^{8+}$   &  $\frac{1}{2}+$ &  2.6289        &  2.8976
     &  1.00651 &  0.00099(3) &  0.00036(18) & -0.00022     &  0.00036     \\
$^{21}$Ne$^{9+}$  &  $\frac{3}{2}+$ & -0.66180(1)    &  2.9672
     &  1.00805 &  0.00113(4) &  0.00058(29) & -0.00037     &  0.00040     \\
$^{23}$Na$^{10+}$ &  $\frac{3}{2}+$ &  2.2175        &  2.9936
     &  1.00975 &  0.00127(4) &  0.00035(18) & -0.00053     &  0.00044     \\
$^{25}$Mg$^{11+}$ &  $\frac{5}{2}+$ & -0.85545(8)    &  3.0280 
     &  1.01163 &  0.00141(4) &  0.00057(29) & -0.00068     &  0.00049     \\
$^{27}$Al$^{12+}$ &  $\frac{5}{2}+$ &  3.6415        &  3.0605        
     &  1.01367 &  0.00156(5) &  0.00048(24) & -0.00084     &  0.00053     \\
$^{29}$Si$^{13+}$ &  $\frac{1}{2}+$ & -0.55529(3)    &  3.1168 
     &  1.01589 &  0.00173(5) &  0.00063(31) & -0.00099     &  0.00058     \\
$^{31}$P$^{14+}$  &  $\frac{1}{2}+$ &  1.1316        &  3.1888 
     &  1.01828 &  0.00191(5) &  0.00069(35) & -0.00115     &  0.00062     \\
$^{33}$S$^{15+}$  &  $\frac{3}{2}+$ &  0.64382       &  3.2727 
     &  1.02085 &  0.00212(6) &  0.00106(53) & -0.00130     &  0.00067     \\
$^{35}$Cl$^{16+}$ &  $\frac{3}{2}+$ &  0.82187       &  3.3652 
     &  1.02360 &  0.00234(6) & -0.00026(110)& -0.00146     &  0.00071     \\
$^{37}$Cl$^{16+}$ &  $\frac{3}{2}+$ &  0.68412       &  3.3840
     &  1.02360 &  0.00235(6) & -0.00055(140)& -0.00146     &  0.00071     \\
$^{39}$K$^{18+}$  &  $\frac{3}{2}+$ &  0.39147       &  3.4346
     &  1.02964 &  0.00274(6) & -0.0021(31)  & -0.00177     &  0.00081     \\
$^{41}$K$^{18+}$  &  $\frac{3}{2}+$ &  0.21487       &  3.4514       
     &  1.02964 &  0.00275(6) & -0.0050(60)  & -0.00177     &  0.00081     \\
$^{43}$Ca$^{19+}$ &  $\frac{7}{2}-$ & -1.3176        &  3.4928
     &  1.03294 &  0.00297(7) &  0.00119(60) & -0.00193     &  0.00086     \\
$^{45}$Sc$^{20+}$ &  $\frac{7}{2}-$ &  4.7565        &  3.5443
     &  1.03644 &  0.00320(7) &  0.00092(46) & -0.00208     &  0.00091     \\
$^{47}$Ti$^{21+}$ &  $\frac{5}{2}-$ & -0.78848(1)    &  3.5944
     &  1.04012 &  0.00345(7) &  0.00160(80) & -0.00224     &  0.00096     \\
$^{49}$Ti$^{21+}$ &  $\frac{7}{2}-$ & -1.1042        &  3.5735
     &  1.04012 &  0.00343(7) &  0.00137(69) & -0.00224     &  0.00096     \\
$^{51}$V$^{22+}$  &  $\frac{7}{2}-$ &  5.1487        &  3.5994
     &  1.04401 &  0.00367(8) &  0.00107(54) & -0.00240     &  0.00101     \\
$^{53}$Cr$^{23+}$ &  $\frac{3}{2}-$ & -0.47454(3)    &  3.6588
     &  1.04810 &  0.00395(8) &  0.00149(75) & -0.00256     &  0.00106     \\
$^{55}$Mn$^{24+}$ &  $\frac{5}{2}-$ &  3.4687        &  3.7057
     &  1.05239 &  0.00423(8) &  0.00109(54) & -0.00272     &  0.00111     \\
$^{57}$Fe$^{25+}$ &  $\frac{1}{2}-$ &  0.090623      &  3.7534 
     &  1.05689 &  0.00453(9) &  0.00279(140)& -0.00289     &  0.00117     \\
$^{59}$Co$^{26+}$ &  $\frac{7}{2}-$ &  4.627(9)      &  3.7875
     &  1.06161 &  0.00483(9) &  0.00133(66) & -0.00305     &  0.00123     \\
$^{61}$Ni$^{27+}$ &  $\frac{3}{2}-$ &  -0.75002(4)   &  3.8221
     &  1.06655 &  0.00514(9) &  0.00191(95) & -0.00322     &  0.00128     \\ \hline
\end{tabular}
\end{table}
\begin{table}
\caption{Comparison of the ground-state
hyperfine splitting $\Delta E^{(1s)}$ of H-like ions between different
theoretical calculations.
The values of the energies are given in meV.}
\label{tab:H:comp}
\tabcolsep10pt
\scriptsize
\begin{tabular}{llccc} \hline
Ion               & $\mu/\mu_N$ 
      & this work   & \cite{shabaev:1995:3686} & \cite{boucard:2000:59} \\ \hline
$^{14}$N$^{6+}$   & 0.40376               
      &  0.21936(6) &  0.21937(4)              &  0.21931               \\
$^{27}$Al$^{12+}$ & 3.6415
      & 10.215(2)   &                          & 10.215                 \\
$^{45}$Sc$^{20+}$ & 4.7565
      & 54.613(25)  &                          & 54.601                 \\
$^{57}$Fe$^{25+}$ & 0.090623
      &  3.5109(49) &                          &  3.5152                \\ \hline
\end{tabular}
\end{table}
\subsection{Li-like ions}
The consideration of Li-like ions we start with
the results for the screened radiative corrections
$x_{\rm SE}$ and $x_{\rm VP}$.
The one-loop QED correction is conveniently
represented in terms of the function $D_{\rm rad}$ defined as
\be
\label{x-D}
  x_{\rm rad} = \frac{\alpha}{\pi}\,D_{\rm rad}\,.
\ee
In Tables~\ref{tab:Li-SE} and \ref{tab:Li-VP}
the numerical results for individual contributions to the
self-energy ($D_{\rm SE}$) and vacuum-polarization ($D_{\rm VP}$)
corrections are presented, respectively,
for $Z = 10,\,15,\,20,\,25$.
\begin{table}
\caption{Individual contributions to the screened self-energy
correction for the ground-state hyperfine structure of the lithiumlike ions,
in units of the function $D_{\rm SE}$.}
\label{tab:Li-SE}
\tabcolsep10pt
\begin{tabular}{rcccc} \hline
$Z$ & $D^{\rm irr}_{\rm SE}$
             & $D^{\rm vr(0)}_{\rm SE}$
                      & $D^{\rm vr(1+)}_{\rm SE}$
                               & $D_{\rm SE}$ \\ \hline
10  & -0.161 &  2.723 & -2.661 & -0.100 \\
15  & -0.296 &  2.685 & -2.760 & -0.372 \\
20  & -0.443 &  2.543 & -2.772 & -0.672 \\
25  & -0.602 &  2.380 & -2.775 & -0.999 \\ \hline
\end{tabular}
\end{table}
\begin{table}
\caption{Individual contributions to the screened vacuum-polarization
correction for the ground-state hyperfine structure of the lithiumlike ions,
in units of the function $D_{\rm VP}$.}
\label{tab:Li-VP}
\tabcolsep10pt
\begin{tabular}{rccllc} \hline
$Z$ & $D^{\rm Ue-el}_{\rm VP}$
             & $D^{\rm Ue-ml}_{\rm VP}$
                      & $D^{\rm WK-el}_{\rm VP}$
                                   & $D^{\rm WK-ml}_{\rm VP}$
                                               & $D_{\rm VP}$ \\ \hline
10  &  0.067 &  0.058 & -0.000037  & -0.00020  & 0.125 \\
15  &  0.119 &  0.098 & -0.00015   & -0.00076  & 0.216 \\
20  &  0.178 &  0.140 & -0.00039   & -0.0020   & 0.316 \\
25  &  0.247 &  0.186 & -0.00083   & -0.0041   & 0.428 \\ \hline
\end{tabular}
\end{table}

Table~\ref{tab:Li} displays the individual contributions to the
hyperfine splitting of the light Li-like ions.
As in the case of H-like ions the main uncertainty
originates from the Bohr-Weisskopf correction $\veps$.
Earlier calculations on the
hyperfine structure of light Li-like ions
\cite{shabaev:1995:3686,shabaev:1997:243,boucard:2000:59}
account for
the radiative correction on the basis of
analytical expansion with respect to $\aZ$.
Here we have performed exact (to all orders in $\aZ$) evaluations
of one-loop QED corrections with an effective
screening potential and with a nuclear vector potential
involving an extended magnetization distribution.
As compared to the results of works
\cite{shabaev:1995:3686,shabaev:1997:243},
several additional improvements have been achieved:
the interelectronic-interaction corrections $B(\aZ)/Z$ and $C(Z,\aZ)/Z^2$
have been calculated taking into account explicitly the extended nuclear charge
and magnetic moment distribution effects, moreover,
the term $C(Z,\aZ)/Z^2$ has now been evaluated within the framework
of the relativistic CI-DFS method.
The latter is especially important for ions of the higher $Z$ region.
The remaining one-electron corrections
$A(\aZ)$, $\delta$, and $\veps$
coincide with the results of works
\cite{shabaev:1995:3686,shabaev:1997:243}.
The recoil effect for Li-like ions
is partly accounted for by a factor $(1+m/M)^{-3}$
in Eq.~(\ref{hfs:1}). However, additional contributions
arising from the specific mass shift and spin-orbit
recoil corrections \cite{yerokhin:2008:012513}
are significantly smaller than the uncertainty assigned
for the Bohr-Weisskopf correction.

In Table~\ref{tab:Li:comp}
results for the total ground-state hyperfine splitting values
of lithiumlike ions $\Delta E^{(2s)}$
of different theoretical calculations
are compared. In addition the experimental value
of a recent measurement
of the hyperfine splitting of lithiumlike $^{45}$Sc$^{18+}$ ion,
performed by resolving the $2s$ hyperfine structure
in the dielectronic recombination spectrum \cite{lestinsky:2008:033001},
is given as well.
The deviations between our results and values
reported in Ref.~\cite{boucard:2000:59}
are mainly determined by interelectronic-interaction effects.
The total theoretical values
of the energies $\Delta E^{(2s)}$ and wavelengths $\lambda^{(2s)}$
are reported in Table~\ref{tab:total}.
\begin{table}
\caption{Individual contributions to the ground-state hyperfine splitting
of the lithiumlike ions.
The values of the nuclear parameters are
the same as in Table~\ref{tab:H}.}
\label{tab:Li}
\tabcolsep10pt
\scriptsize
\begin{tabular}{lccccccc} \hline
Ion
     & $A(\aZ)$ & $\delta$    & $\veps$      & $x_{\rm SE}$ 
  & $x_{\rm VP}$ & $B(\aZ)/Z$  & $C(Z,\aZ)/Z^2$ \\ \hline
$^{14}$N$^{4+}$
     &  1.00557 &  0.00067(3) & -0.00004(28) &  0.00008 
  &  0.00017     & -0.38146    &  0.01800     \\
$^{15}$N$^{4+}$
     &  1.00557 &  0.00068(3) &  0.00114(88) &  0.00008  
  &  0.00017     & -0.38101    &  0.01798     \\
$^{17}$O$^{5+}$
     &  1.00728 &  0.00081(3) &  0.00033(17) & -0.00001  
  &  0.00021     & -0.33424    &  0.01387     \\
$^{19}$F$^{6+}$
     &  1.00923 &  0.00099(3) &  0.00036(18) & -0.00012   
  &  0.00025     & -0.29767    &  0.01104     \\
$^{21}$Ne$^{7+}$
     &  1.01142 &  0.00113(4) &  0.00058(29) & -0.00023  
  &  0.00029     & -0.26845    &  0.00900     \\
$^{23}$Na$^{8+}$
     &  1.01384 &  0.00127(4) &  0.00035(18) & -0.00035 
  &  0.00033     & -0.24471    &  0.00749     \\
$^{25}$Mg$^{9+}$
     &  1.01650 &  0.00141(4) &  0.00057(29) & -0.00047 
  &  0.00037     & -0.22488    &  0.00634     \\
$^{27}$Al$^{10+}$
     &  1.01941 &  0.00156(5) &  0.00048(24) & -0.00060  
  &  0.00042     & -0.20823    &  0.00544     \\
$^{29}$Si$^{11+}$
     &  1.02257 &  0.00173(5) &  0.00063(32) & -0.00073  
  &  0.00046     & -0.19395    &  0.00473     \\
$^{31}$P$^{12+}$
     &  1.02597 &  0.00192(5) &  0.00070(35) & -0.00086    
  &  0.00050     & -0.18164    &  0.00415     \\
$^{33}$S$^{13+}$
     &  1.02963 &  0.00212(6) &  0.00107(53) & -0.00100   
  &  0.00055     & -0.17086    &  0.00368     \\
$^{35}$Cl$^{14+}$
     &  1.03355 &  0.00235(6) & -0.00026(111)& -0.00113   
  &  0.00059     & -0.16166    &  0.00329     \\
$^{37}$Cl$^{14+}$
     &  1.03355 &  0.00236(6) & -0.00055(140)& -0.00113  
  &  0.00059     & -0.16171    &  0.00329     \\
$^{39}$K$^{16+}$
     &  1.04219 &  0.00275(6) & -0.0021(31)  & -0.00142 
  &  0.00069     & -0.14619    &  0.00268     \\
$^{41}$K$^{16+}$
     &  1.04219 &  0.00276(6) & -0.0050(60)  & -0.00142         
  &  0.00069     & -0.14661    &  0.00269     \\
$^{43}$Ca$^{17+}$
     &  1.04691 &  0.00298(7) &  0.00120(60) & -0.00156 
  &  0.00073     & -0.13908    &  0.00244     \\
$^{45}$Sc$^{18+}$
     &  1.05191 &  0.00322(7) &  0.00092(46) & -0.00171 
  &  0.00078     & -0.13316    &  0.00223     \\
$^{47}$Ti$^{19+}$
     &  1.05719 &  0.00347(7) &  0.00161(81) & -0.00186  
  &  0.00083     & -0.12769    &  0.00205     \\
$^{49}$Ti$^{19+}$
     &  1.05719 &  0.00345(7) &  0.00138(69) & -0.00186  
  &  0.00083     & -0.12772    &  0.00205     \\
$^{51}$V$^{20+}$
     &  1.06277 &  0.00369(8) &  0.00108(54) & -0.00201 
  &  0.00089     & -0.12288    &  0.00190     \\
$^{53}$Cr$^{21+}$
     &  1.06864 &  0.00398(8) &  0.00151(75) & -0.00216
  &  0.00094     & -0.11840    &  0.00176     \\
$^{55}$Mn$^{22+}$
     &  1.07481 &  0.00427(8) &  0.00110(55) & -0.00232 
  &  0.00099     & -0.11440    &  0.00164     \\
$^{57}$Fe$^{23+}$
     &  1.08130 &  0.00457(9) &  0.00282(141)& -0.00248  
  &  0.00105     & -0.11050    &  0.00154     \\
$^{59}$Co$^{24+}$
     &  1.08811 &  0.00487(9) &  0.00134(67) & -0.00264
  &  0.00111     & -0.10728    &  0.00144     \\
$^{61}$Ni$^{25+}$
     &  1.09524 &  0.00519(9) &  0.00193(96) & -0.00281
  &  0.00117     & -0.10410    &  0.00136     \\ \hline
\end{tabular}
\end{table}
\begin{table}
\caption{Comparison of the ground-state
hyperfine splitting $\Delta E^{(2s)}$ of Li-like ions between different
theoretical calculations and experimental data.
The values of the energies are given in meV.}
\label{tab:Li:comp}
\tabcolsep10pt
\scriptsize
\begin{tabular}{llllll} \hline
Ion               & $\mu/\mu_N$ 
      & this work   & \cite{shabaev:1997:243}  & \cite{boucard:2000:59} 
  & Exp. \cite{lestinsky:2008:033001} \\ \hline
$^{14}$N$^{4+}$   & 0.40376               
      & 0.017532(8) & 0.017532(10)             &  0.017667
  &                                \\
$^{27}$Al$^{10+}$ & 3.6415
      & 1.0283(3)   & 1.0281(6)                &  1.0326
  &                                \\
$^{45}$Sc$^{18+}$ & 4.7565
      & 6.0631(32)  & 6.063(6)                 &  6.0767
  & 6.20(8)                        \\
$^{57}$Fe$^{23+}$ & 0.090623
      & 0.40345(64) & 0.4036(7)                &  0.40470
  &                                \\ \hline
\end{tabular}
\end{table}
\subsection{B-like ions}
Let us now turn to boronlike ions.
The corresponding screened radiative corrections $x_{\rm SE}$ and $x_{\rm VP}$
expressed in terms of the functions $D_{\rm SE}$ and $D_{\rm VP}$
defined by Eq.~(\ref{x-D})
are presented in Tables~\ref{tab:B-SE} and \ref{tab:B-VP},
respectively.
The uncalculated Wichmann-Kroll contribution of the magnetic-loop
$D^{\rm WK-ml}_{\rm VP}$ is of the same order
as the corresponding electric-loop term
$D^{\rm WK-el}_{\rm VP}$. As can be seen from Table~\ref{tab:B-VP},
the values of the correction $D^{\rm WK-el}_{\rm VP}$
are smaller than the uncertainty of the self-energy contribution
$D_{\rm SE}$. Therefore, in the case of light B-like ions one
can neglect the contributions of the Wichmann-Kroll terms.
\begin{table}
\caption{Individual contributions to the screened self-energy
correction for the ground-state hyperfine splitting of the boronlike ions,
in units of the function $D_{\rm SE}$.}
\label{tab:B-SE}
\tabcolsep10pt
\begin{tabular}{rcccc} \hline
$Z$ & $D^{\rm irr}_{\rm SE}$
             & $D^{\rm vr(0)}_{\rm SE}$
                      & $D^{\rm vr(1+)}_{\rm SE}$
                               &  $D_{\rm SE}$ \\ \hline
10  &  0.000 &  0.630 & -0.521 &  0.109 \\
15  & -0.001 &  0.742 & -0.608 &  0.132 \\
20  & -0.002 &  0.781 & -0.648 &  0.131 \\
25  & -0.005 &  0.788 & -0.666 &  0.117 \\ \hline
\end{tabular}
\end{table}
\begin{table}
\caption{Individual contributions to the screened vacuum-polarization
correction for the ground-state hyperfine splitting of the boronlike ions,
in units of the function $D_{\rm VP}$.}
\label{tab:B-VP}
\tabcolsep10pt
\begin{tabular}{rllll} \hline
$Z$ & $D^{\rm Ue-el}_{\rm VP}$
               & $D^{\rm Ue-ml}_{\rm VP}$
                          & $D^{\rm WK-el}_{\rm VP}$
                                            & $D_{\rm VP}$ \\ \hline
10  &  0.00011 &  0.00066 & -0.000000084    & 0.00078 \\
15  &  0.00059 &  0.0023  & -0.00000097     & 0.0029  \\
20  &  0.0018  &  0.0052  & -0.0000051      & 0.0069  \\
25  &  0.0041  &  0.0096  & -0.000018       & 0.0137  \\ \hline
\end{tabular}
\end{table}

In Table~\ref{tab:B} numerical results for the
individual contributions to the hyperfine splitting
in light B-like ions are displayed.
In contrast to the hydrogen- and lithiumlike sequences,
where the valence electrons are in the $s$ states,
boronlike ions have the valence electron
in the $p_{1/2}$ state; its electron
density vanishes at the origin.
Since the nuclear structure corrections $\delta$
and $\veps$ arise from the nuclear region,
these contributions are much smaller in
the B-like ions than in corresponding H- and Li-like ions.
For the uncertainty of the radiative correction
we prefer a conservative estimation as
the difference of QED corrections calculated with and without
screening potential.
This uncertainty dominates for ions in the low-$Z$ region.
Evaluation of the screened radiative correction within the
rigorous QED approach is presently underway.
For high-$Z$ ions the total theoretical uncertainty
is mainly determined by the frequency-dependent (QED) contribution
in the higher-order interelectronic-interaction term $C(Z,\aZ)/Z^2$,
which is estimated to be of the order $(\aZ)^3 C(Z,\aZ)/Z^2$.
In Table~\ref{tab:total} the total theoretical values
of the energies $\Delta E^{(2p_{1/2})}$
and wavelengths $\lambda^{(2p_{1/2})}$ are reported.
The recoil correction is accounted for by a factor $(1+m/M)^{-3}$
in Eq.~(\ref{hfs:1}) with 100\% uncertainty, caused by uncalculated
specific mass shift and spin-orbit recoil corrections
\cite{yerokhin:2008:012513}.
\begin{table}
\caption{Individual contributions to the ground-state hyperfine splitting
of the boronlike ions.
The values of the nuclear parameters are
the same as in Table~\ref{tab:H}.}
\label{tab:B}
\tabcolsep10pt
\scriptsize
\begin{tabular}{lcllcll} \hline
Ion
    & $A(\aZ)$ &\;\;$\delta$    &\;\;$\veps$    & $x_{\rm rad}$ 
  &$B(\aZ)/Z$ &$C(Z,\aZ)/Z^2$ \\ \hline
$^{14}$N$^{2+}$
    &  1.00513 &  0.000001      & 0.000000(1)   &  0.00017(37)
  & -0.82166    &  0.15006(2)    \\
$^{15}$N$^{2+}$
    &  1.00513 &  0.000001      & 0.000002(2)   &  0.00017(37)
  & -0.82166    &  0.15006(2)    \\
$^{17}$O$^{3+}$
    &  1.00671 &  0.000002      & 0.000001      &  0.00020(33)
  & -0.72061    &  0.11479(2)    \\
$^{19}$F$^{4+}$   
    &  1.00850 &  0.000002      & 0.000001(1)   &  0.00023(29)
  & -0.64222    &  0.09083(3)    \\
$^{21}$Ne$^{5+}$
    &  1.01052 &  0.000003      & 0.000002(1)   &  0.00025(26)
  & -0.57969    &  0.07381(3)    \\
$^{23}$Na$^{6+}$
    &  1.01275 &  0.000005      & 0.000002(1)   &  0.00027(24)
  & -0.52869    &  0.06128(3)    \\
$^{25}$Mg$^{7+}$
    &  1.01520 &  0.000006      & 0.000003(1)   &  0.00029(21)
  & -0.48635    &  0.05179(3)    \\
$^{27}$Al$^{8+}$
    &  1.01788 &  0.000008      & 0.000003(1)   &  0.00030(19)
  & -0.45068    &  0.04443(4)    \\
$^{29}$Si$^{9+}$
    &  1.02078 &  0.000010      & 0.000004(2)   &  0.00031(17)
  & -0.42023    &  0.03860(4)    \\
$^{31}$P$^{10+}$
    &  1.02392 &  0.000013      & 0.000005(3)   &  0.00031(16)
  & -0.39398    &  0.03391(4)    \\
$^{33}$S$^{11+}$
    &  1.02728 &  0.000016      & 0.000009(5)   &  0.00032(14)
  & -0.37113    &  0.03008(5)    \\
$^{35}$Cl$^{12+}$
    &  1.03089 &  0.000021(1)   &-0.000003(11)  &  0.00032(13)
  & -0.35110    &  0.02691(5)    \\
$^{37}$Cl$^{12+}$
    &  1.03089 &  0.000021(1)   &-0.000006(14)  &  0.00032(13)
  & -0.35110    &  0.02691(5)    \\
$^{39}$K$^{14+}$
    &  1.03882 &  0.000030(1)   &-0.000026(38)  &  0.00032(10)
  & -0.31768    &  0.02203(6)    \\
$^{41}$K$^{14+}$
    &  1.03882 &  0.000030(1)   &-0.000062(74)  &  0.00032(10)
  & -0.31770    &  0.02203(6)    \\
$^{43}$Ca$^{15+}$
    &  1.04316 &  0.000036(1)   & 0.000016(8)   &  0.00032(9)
  & -0.30363    &  0.02012(6)    \\
$^{45}$Sc$^{16+}$
    &  1.04775 &  0.000043(1)   & 0.000014(7)   &  0.00032(8)
  & -0.29103    &  0.01849(7)    \\
$^{47}$Ti$^{17+}$
    &  1.05260 &  0.000051(1)   & 0.000026(13)  &  0.00032(7)
  & -0.27967    &  0.01708(7)    \\
$^{49}$Ti$^{17+}$
    &  1.05260 &  0.000051(1)   & 0.000023(11)  &  0.00032(7)
  & -0.27967    &  0.01708(7)    \\
$^{51}$V$^{18+}$
    &  1.05772 &  0.000060(1)   & 0.000019(10)  &  0.00031(6)
  & -0.26941    &  0.01585(7)    \\
$^{53}$Cr$^{19+}$
    &  1.06311 &  0.000070(1)   & 0.000029(15)  &  0.00031(5)
  & -0.26010    &  0.01477(8)    \\
$^{55}$Mn$^{20+}$
    &  1.06877 &  0.000082(2)   & 0.000023(12)  &  0.00030(5)
  & -0.25163    &  0.01383(8)    \\
$^{57}$Fe$^{21+}$
    &  1.07472 &  0.000095(2)   & 0.000064(32)  &  0.00030(4)
  & -0.24390(1) &  0.01300(9)    \\
$^{59}$Co$^{22+}$
    &  1.08096 &  0.000109(2)   & 0.000033(16)  &  0.00029(3)
  & -0.23685(1) &  0.01226(9)    \\
$^{61}$Ni$^{23+}$
    &  1.08749 &  0.000125(2)   & 0.000051(25)  &  0.00029(2)
  & -0.23039(1) &  0.01160(10)   \\ \hline
\end{tabular}
\end{table}
%
%%%%%%%%%%%%%%%%%%%%%%%%%%%%%%%%%%%%%%%%%%%%%%%%%%%%%%%%%%%%%%%%%%%%%%%%
%
\section{Discussion}
\label{sec-4}
Predictions for the total energies $\Delta E^{(a)}$
and wavelengths $\lambda^{(a)}$
of the transitions between the ground-state hyperfine splitting
components of the light H-, Li-, and B-like ions are given
in Table~\ref{tab:total}.
Due to the discrepancies in the experimental
data for nuclear magnetic moments $\mu$
for some ions we have evaluated transition energies and wavelengths
for all values of $\mu$ reported in Ref.~\cite{stone:2005:75}.
The values of the nuclear spin and parity,
and the root-mean-square radii are
the same as in Table~\ref{tab:H}.
In the parentheses the uncertainty of the presented results
is indicated.
For $1s$ and $2s$ states it is mainly
due to the Bohr-Weisskopf effect and must generally be
considered as the order of
magnitude of the expected error bar.
For some ions we give also a second
value for the uncertainty, which corresponds
to the uncertainty of the nuclear magnetic moment.
The values for lithiumlike $^{45}$Sc$^{18+}$
and boronlike $^{45}$Sc$^{16+}$, $^{57}$Fe$^{21+}$
ions coincide with our previous results
\cite{oreshkina:2007:889,kozhedub:2007:012511,oreshkina:2008:675}.
Due to the lack of experimental data
one can not make a detailed comparison for the
ions under consideration.

\begin{table}
\caption{The energies $\Delta E^{(a)}$ (meV) and wavelengths $\lambda^{(a)}$ (cm)
of the transitions between the ground-state hyperfine splitting
of the H-, Li-, and B-like ions.}
\label{tab:total}
\tabcolsep6pt
\scriptsize
\begin{tabular}{llllllll} \hline
Nucleus   &  $\mu/\mu_N$
   & $\Delta E^{(1s)}$ & $\lambda^{(1s)}$ 
       & $\Delta E^{(2s)}$ & $\lambda^{(2s)}$ 
          & $\Delta E^{(2p_{1/2})}$ & $\lambda^{(2p_{1/2})}$ \\ \hline
$^{14}$N   &   0.40376
   &   0.21936(6)      & 0.56521(16)
       &   0.017532(8)     & 7.0720(31)
          &  0.0030389(34)          & 40.799(46)       \\
$^{15}$N   &  -0.28319
   &   0.20490(18)     & 0.60510(53)    
       &   0.016376(23)    & 7.5711(105)
          &  0.0028419(32)          & 43.627(49)       \\
$^{17}$O   &  -1.8938(1)
   &   1.2294(2)(1)    & 0.10085(2)(1)
       &   0.10497(3)(1)   & 1.1811(3)(1)
          &   0.020459(17)(1)       &  6.0602(50)(3)   \\
$^{19}$F   &   2.6289
   &   4.0541(7)       & 0.030582(6) 
       &   0.36364(9)      & 0.34095(9)
          &   0.076848(49)          &  1.6134(10)      \\
$^{21}$Ne  &  -0.66180(1)
   &   0.93432(27)(1)  & 0.13270(4) 
       &   0.087074(34)(1) & 1.4239(6)
          &   0.019531(10)          &  6.3481(33)(1)   \\
$^{23}$Na  &   2.2175
   &   4.1738(8)       & 0.029705(5) 
       &   0.40110(9)      & 0.30911(7)
          &   0.094130(42)          &  1.3172(6)       \\
           &   2.2177
   &   4.1742(8)       & 0.029703(5)  
       &   0.40113(9)      & 0.30909(7)
          &   0.094138(42)          &  1.3170(6)       \\
$^{25}$Mg  &  -0.85545(8)
   &   1.8840(5)(2)    & 0.065809(19)(6)
       &   0.18567(7)(2)   & 0.66777(25)(6)
          &   0.045176(17)(4)       &  2.7445(10)(3)   \\
$^{27}$Al  &   3.6415
   &  10.215(2)        & 0.012137(3)  
       &   1.0283(3)       & 0.12058(4)
          &   0.25755(8)            &  0.48140(16)     \\
$^{29}$Si  &  -0.55529(3)
   &   3.2484(10)(2)   & 0.038168(12)(2)
       &   0.33293(13)(2)  & 0.37240(15)(2)
          &   0.085431(24)(5)       &  1.4513(4)(1)    \\
$^{31}$P   &   1.1316
   &   8.1582(29)      & 0.015198(5) 
       &   0.84933(36)     & 0.14598(6)
          &   0.22240(6)            &  0.55748(14)     \\
$^{33}$S   &   0.64382
   &   3.7624(20)      & 0.032954(18)
       &   0.39711(25)     & 0.31221(20)
          &   0.10583(2)            &  1.1716(3)       \\
$^{35}$Cl  &   0.82187
   &   5.7821(64)      & 0.021443(24) 
       &   0.61780(81)     & 0.20069(26)
          &   0.16687(3)            &  0.74298(15)     \\
$^{37}$Cl  &   0.68412
   &   4.8143(67)      & 0.025753(36)
       &   0.51439(85)     & 0.24103(40)
          &   0.13891(3)            &  0.89258(18)     \\
$^{39}$K   &   0.39147
   &   3.873(12)       & 0.03202(10)
       &   0.4225(15)      & 0.2934(11)
          &   0.11670(2)            &  1.0625(2)       \\
           &   0.39151
   &   3.873(12)       & 0.03201(10)
       &   0.4226(15)      & 0.2934(11)
          &   0.11671(2)            &  1.0624(2)       \\
$^{41}$K   &   0.21487
   &   2.132(13)       & 0.05816(35)
       &   0.2326(16)      & 0.5331(37)
          &   0.064053(12)          &  1.9356(4)       \\
           &   0.21489
   &   2.132(13)       & 0.05816(35)
       &   0.2326(16)      & 0.5330(37)
          &   0.064059(12)          &  1.9355(4)       \\
$^{43}$Ca  &  -1.3176
   &  13.025(8)        & 0.0095191(57)
       &   1.4340(10)      & 0.086460(60)
          &   0.40134(6)            &  0.30892(5)      \\
$^{45}$Sc  &   4.7565
   &  54.613(25)       & 0.0022702(11)
       &   6.0631(32)      & 0.020449(11)
          &   1.7115(2)             &  0.072440(10)    \\
$^{47}$Ti  &  -0.78848(1)
   &  10.957(9)        & 0.011316(9)
       &   1.2259(11)      & 0.10114(9)
          &   0.34905(5)            &  0.35520(5)      \\
$^{49}$Ti  &  -1.1042
   &  14.617(10)       & 0.0084821(59)
       &   1.6355(13)      & 0.075810(60)
          &   0.46554(6)            &  0.26632(3)      \\
$^{51}$V   &   5.1487
   &  78.168(42)       & 0.0015861(9)
       &   8.8099(54)      & 0.014073(9)
          &   2.5248(3)             &  0.049107(6)     \\
$^{53}$Cr  &  -0.47454(3)
   &   9.5797(72)(6)   & 0.012942(10)(1)
       &   1.0871(9)(1)    & 0.11406(10)(1)
          &   0.31367(4)(2)         &  0.39527(5)(2)   \\
$^{55}$Mn  &   3.4687
   &  71.525(39)       & 0.0017334(9)
       &   8.1688(50)      & 0.015178(9)
          &   2.3699(3)             &  0.052315(6)     \\
           &   3.4532(13)
   &  71.206(39)(27)   & 0.0017412(9)(7)
       &   8.1323(50)(31)  & 0.015246(9)(6)
          &   2.3594(3)(9)          &  0.052550(6)(20) \\
$^{57}$Fe  &   0.090623
   &   3.5109(49)      & 0.035314(50) 
       &   0.40345(64)     & 0.30731(49)
          &   0.11787(1)            &  1.0519(1)       \\
           &   0.090764
   &   3.5164(49)      & 0.035259(50) 
       &   0.40408(64)     & 0.30683(48)
          &   0.11805(1)            &  1.0503(1)       \\
           &   0.09044(7)
   &   3.5038(49)(27)  & 0.035386(50)(27)
       &   0.40264(64)(31) & 0.30793(49)(24)
          &   0.11763(1)(9)         &  1.0540(1)(8)    \\
$^{59}$Co  &   4.627(9)
   & 115.35(8)(22)     & 0.0010749(7)(21)
       &  13.334(10)(26)   & 0.0092987(69)(181)
          &   3.9084(5)(76)         &  0.031722(4)(62) \\
$^{61}$Ni  &  -0.75002(4)
   &  24.418(23)(1)    & 0.0050777(49)(3)
       &   2.8385(30)(2)   & 0.043680(47)(2)
          &   0.83617(10)(4)        &  0.14828(2)(1)   \\ \hline
\end{tabular}
\end{table}
Table~\ref{tab:total} shows excellent accuracy for the values of the
hyperfine splitting of H-, Li- and B-like ions.
For the H- and Li-like ions the limitation of the total accuracy
is set by the Bohr-Weisskopf correction.
However, this uncertainty can be considerably
reduced in the specific difference
of the ground-state hyperfine structure values of H- and Li-like ions
with the same nucleus
\cite{shabaev:2001:3959}
\be
\label{specdiff}
  \Delta^\prime E = \Delta E^{(2s)} - \xi \Delta E^{(1s)}\,.
\ee
The parameter $\xi$ has to be chosen to cancel
the Bohr-Weisskopf correction
\be
\label{xi}
  \xi = \frac{1}{8}\,
  \frac{A^{(2s)}(\aZ)(1-\delta^{(2s)})}{A^{(1s)}(\aZ)(1-\delta^{(1s)})}
  \,f(\aZ)\,.
\ee
The function $f(\aZ)$ is defined by the ratio of the Bohr-Weisskopf
corrections 
\be
  f(\aZ) = \frac{\veps^{(2s)}}{\veps^{(1s)}}\,.
\ee
This function can be calculated to a rather high accuracy,
because it is determined mainly by the
behavior of the wavefunctions at the atomic scale
and thus almost independent of the nuclear
structure \cite{shabaev:2001:3959}.
In Table~\ref{tab:xi} we present the numerical results
for the parameter $\xi$ and for the specific difference $\Delta^\prime E$
between the hyperfine splitting values of H- and Li-like ions.
The theoretical accuracy of the presented values $\Delta^\prime E$
is better than $0.01\%$ and the given uncertainty is determined
by the limited knowledge of the nuclear magnetic moments.
\begin{table}
\caption{The parameter $\xi$ and specific difference $\Delta^\prime E$ (meV)
between the hyperfine structure values of H- and Li-like
ions, defined by Eqs.~(\ref{xi}) and (\ref{specdiff}), respectively.}
\label{tab:xi}
\tabcolsep6pt
\scriptsize
\begin{tabular}{llll} \hline
Nucleus    &  $\mu/\mu_N$ & $\xi$   & $\Delta^\prime E$ \\ \hline
$^{14}$N   &   0.40376    & 0.12531 & -0.009955    \\
$^{15}$N   &  -0.28319    & 0.12530 & -0.009298    \\
$^{17}$O   &  -1.8938(1)  & 0.12539 & -0.04919     \\
$^{19}$F   &   2.6289     & 0.12550 & -0.1451      \\
$^{21}$Ne  &  -0.66180(1) & 0.12561 & -0.03029     \\
$^{23}$Na  &   2.2175     & 0.12574 & -0.1237      \\
           &   2.2177     & 0.12574 & -0.1237      \\
$^{25}$Mg  &  -0.85545(8) & 0.12588 & -0.05148     \\
$^{27}$Al  &   3.6415     & 0.12603 & -0.2592      \\
$^{29}$Si  &  -0.55529(3) & 0.12619 & -0.07699     \\
$^{31}$P   &   1.1316     & 0.12637 & -0.1816      \\
$^{33}$S   &   0.64382    & 0.12656 & -0.07904     \\
$^{35}$Cl  &   0.82187    & 0.12676 & -0.1151      \\
$^{37}$Cl  &   0.68412    & 0.12676 & -0.09587     \\
$^{39}$K   &   0.39147    & 0.12720 & -0.07006     \\
           &   0.39151    & 0.12720 & -0.07007     \\
$^{41}$K   &   0.21487    & 0.12720 & -0.03856     \\
           &   0.21489    & 0.12720 & -0.03857     \\
$^{43}$Ca  &  -1.3176     & 0.12743 & -0.2258      \\
$^{45}$Sc  &   4.7565     & 0.12768 & -0.9098      \\
$^{47}$Ti  &  -0.78848(1) & 0.12794 & -0.1759      \\
$^{49}$Ti  &  -1.1042     & 0.12794 & -0.2347      \\
$^{51}$V   &   5.1487     & 0.12822 & -1.213       \\
$^{53}$Cr  &  -0.47454(3) & 0.12851 & -0.1440      \\
$^{55}$Mn  &   3.4687     & 0.12881 & -1.044       \\
           &   3.4532(13) & 0.12881 & -1.039       \\
$^{57}$Fe  &   0.090623   & 0.12912 & -0.04989     \\
           &   0.090764   & 0.12912 & -0.04996     \\
           &   0.09044(7) & 0.12912 & -0.04979(4)  \\
$^{59}$Co  &   4.627(9)   & 0.12945 & -1.599(3)    \\
$^{61}$Ni  &  -0.75002(4) & 0.12979 & -0.3307      \\ \hline
\end{tabular}
\end{table}

Let us summarize: {\it Ab initio} QED calculations of the ground-state hyperfine
splitting of H-, Li-, and B-like ions in the
middle-$Z$ region have been performed.
The evaluation incorporates results based on a rigorous treatment
of first-order many-electron QED effects and
on the large-scale CI-DFS calculations of the second- and higher-order
electron-correlation effects.
The one-loop radiative corrections have been evaluated
to all orders in $\aZ$. The screening QED effect in Li- and B-like
sequences have been taken into account utilizing a
local Kohn-Sham potential. The Bohr-Weisskopf correction has been
calculated employing the single-particle nuclear model.
As the result, the most accurate values for the
hyperfine splitting of H-, Li-, and B-like ions under consideration
have been obtained.
The specific difference of the hyperfine splitting values of H- and
Li-like ions, where the uncertainties associated with
the nuclear-structure corrections are significantly canceled,
has been evaluated with an accuracy better than $0.01\%$.
%
%%%%%%%%%%%%%%%%%%%%%%%%%%%%%%%%%%%%%%%%%%%%%%%%%%%%%%%%%%%%%%%%%%%%%%%%
\acknowledgments
%%%%%%%%%%%%%%%%%%%%%%%%%%%%%%%%%%%%%%%%%%%%%%%%%%%%%%%%%%%%%%%%%%%%%%%%
%
Stimulating discussions with R. A. Sunyaev are gratefully acknowledged.
We also thank A. N. Artemyev for providing us with
the computer code for the evaluation of the Wichmann-Kroll
electric-loop potential.
The authors acknowledge major financial
support from DFG, GSI,
RFBR (Grant No. 07-02-00126a), and
INTAS-GSI (Grant No. 06-1000012-8881).
The work of D.A.G. was also supported by
the grant of President of Russian Federation
(Grant No. MK-3957.2008.2).
N.S.O. acknowledges the support from St. Petersburg Government
(Grant No. 75-MU).
%
%%%%%%%%%%%%%%%%%%%%%%%%%%%%%%%%%%%%%%%%%%%%%%%%%%%%%%%%%%%%%%%%%%%%%%%%
%
%%%%%%%%%%%%%%%%%%%%%%%%%%%%%%%%%%%%%%%%%%%%%%%%%%%%%%%%%%%%%%%%%%%%%%%%

%%%%%%%%%%%%%%%%%%%%%%%%%%%%%%%%%%%%%%%%%%%%%%%%%%%%%%%%%%%%%%%%%%%%%%%%
%
%%%%%%%%%%%%%%%%%%%%%%%%%%%%%%%%%%%%%%%%%%%%%%%%%%%%%%%%%%%%%%%%%%%%%%%%
\end{document}